\documentclass[secnumarabic%
,amssymb, amsmath,nobibnotes, preprint, floatfix]{revtex4-1}

\usepackage{bm}
\usepackage{color}

\usepackage{comment}
\usepackage{extraarrows}
\usepackage{graphicx}

\usepackage{bm}

\def\bal#1\eal{\begin{align}#1\end{align}}

\usepackage{amssymb,amsfonts,amsmath}

\makeatletter 
\def\@cite#1{(#1)}
\makeatother

\def\A{{\bf A}}
\def\m{{\mathfrak m}}
\def\r{{\mathfrak r}}
\def\G{\Gamma}

\begin{document}

\newcommand{\tabenv}[1][\linewidth]{\def\@captype{table}}

\title{Law of Localization in Chemical Reaction Networks}

\author{Takashi Okada$^1$  and Atsushi Mochizuki$^{1,2}$}
\affiliation{%
$^1$Theoretical Biology Laboratory, RIKEN, Wako 351-0198, Japan \\
$^2$CREST, JST
4-1-8 Honcho, Kawaguchi 332-0012, Japan
}



\begin{abstract}

In living cells, chemical reactions are connected by sharing their products and substrates, and form complex networks, e.g. metabolic pathways. Here we developed a theory to predict the
sensitivity, i.e. the responses of concentrations and fluxes to
perturbations of enzymes, from network structure alone. Responses turn
out to exhibit two characteristic patterns, $localization$ and
 $hierarchy$. We present a general theorem connecting sensitivity with
network topology that explains these characteristic patterns. Our
results imply that network topology is an origin of biological
robustness. Finally, we suggest a strategy to determine real networks
from experimental measurements.


\end{abstract}

   
\maketitle


\section{Introduction}

Cells have many  chemical reactions, each of which is mediated by  organic catalysts, enzymes. Reactions are not independent but connected and form complex networks. The dynamics of chemical concentrations are considered as  the origin of physiological functions. However, dynamical behavior based on the network is not understood well. 

\begin{figure}[h] 
  \includegraphics[width=8.0cm,bb=0 10 300 75]{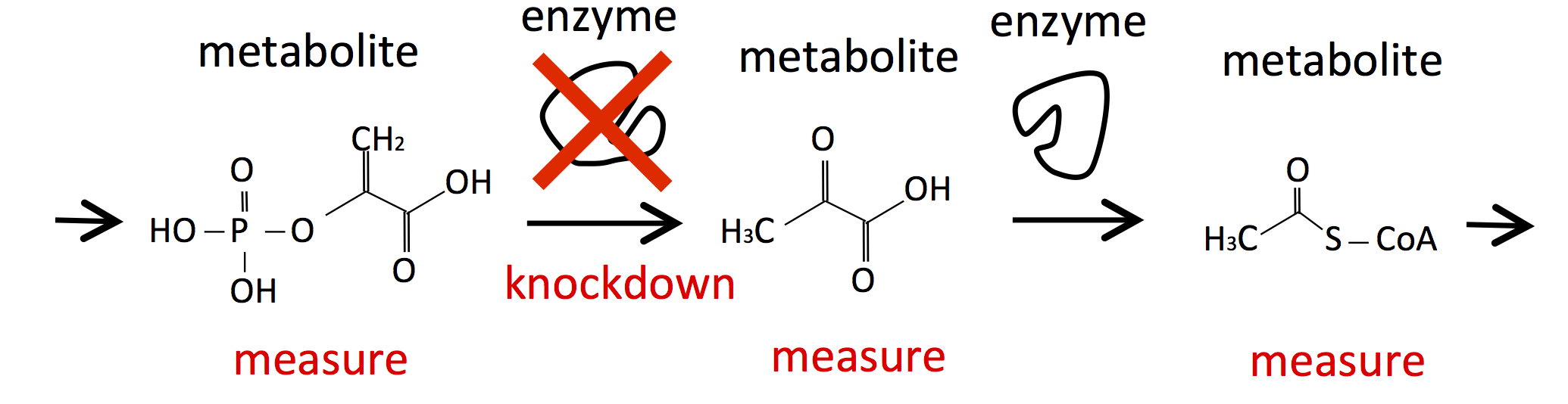}
   \caption{ Sensitivity analysis. After the amount/activity of an  enzyme protein is decreased,  the concentration change of  metabolites are measured.}
\label{fig:KO}
\end{figure}

One experimental approach to study such  network systems is sensitivity analysis where the amount or activity of enzyme is perturbed and responses  (concentrations of chemicals in the system) are measured \cite{Ishii} (see Fig. \ref{fig:KO}). However, the results of such experiments are very difficult to interpret, because theoretical criteria to evaluate the results of perturbations from  network structures are not established.

There are  other difficulties in understanding dynamical behaviors of  reaction systems in biology. First, although a huge amount of  information of reaction networks is available on databases \cite{KEGG,Reactome,BioCyc}, they provide no more than knowledge of identified reactions in  biochemistry. It is possible that the information is incomplete, including many unidentified reactions or regulations.  Second, 
 in spite of the recent progress in biosciences, it is still difficult or almost impossible to  determine   quantitative details of the dynamics, such as  functions for  reaction rates,  parameter values, or initial states. 


In order to circumvent these difficulties, we introduce a mathematical method, named {\it structural sensitivity analysis} \cite{Mochizuki, monomolecular}, to determine responses of chemical reaction systems to the perturbation of the enzyme amount or activity based only on network structure. From analyses we found that qualitative responses  at a steady state are determined from information of network structure only. We also found that response patterns, e.g. distribution of nonzero responses of chemical concentrations in the network,  exhibit two characteristic features,  $localization$ and $hierarchy$ depending on the structure of networks and position of perturbed reactions. Finally we found a general theorem connecting the network topology and the response patterns directly, and governing the characteristic patterns of responses. This theorem, which we call  the {\it law of localization}, is not only theoretically important, but also practically useful for examining  real biological systems.
\textcolor{black}{In the context of adaptation, there were some previous studies, which reported confined nonzero responses in specific  systems \cite{
Ni,Steuer,Drengstig,He}. However, they did not find general laws of such response patterns, nor any topological conditions.}

\section{Structural Sensitivity Analysis}
We study concentration changes in a reaction system under perturbation of reaction rate parameters, assuming that the system is in a steady state \cite{Mochizuki, monomolecular}. We label chemicals by  $m \,(m=1,\ldots,M)$ and reactions by  $j\,(j=1,\ldots,R)$. A state of the system is specified by  concentrations $x_m (t)$ and  obeys the following differential equations \cite{F1,F2}
\begin{align}
\frac{d x_m}{dt}= \sum_{i=1}^R { {\bm \nu}}_{mi} W_i(k_i;x). \label{ode}
\end{align}
Here,  $\bm \nu$ is called a stoichiometric matrix.  $W_i$ is called a flux, which depends on metabolite concentrations and also on a reaction rate  $k_i$.  We do not assume specific forms for $W_i$, but assume that each $W_i$ is an increasing function of its substrate concentration;
\begin{align}
\begin{cases}
 \frac{\partial W_i}{\partial x_m}     > 0 \  \    {\rm  if} \ x_m {\rm \ is \ a \ substrate \  of \ reaction \ } i, \\
  \frac{\partial W_i}{\partial x_m} =0\  \  {\rm otherwise.}\ 
  \end{cases}\label{noreg}
\end{align}
Below,  we abbreviate and emphasize nonzero $\frac{\partial W_i}{\partial x_m}$  as $r_{im}$. 


In this framework, enzyme knockdown of  the $j$-th reaction corresponds to changing the rate as $k_j \rightarrow k_j+ \delta k_j$ \textcolor{black}{(triangles in   FIG.    \ref{fig:NW})}. By assuming steady state \cite{Palsson, FBA1,FBA2,FBA3}, the flux is expressed, in terms of  a basis $\{\vec c_n\}$ of  ${\rm ker}\, {\bm \nu}$, as $\vec W =\sum_{n=1}^{N_k}\, \mu^n\,  \vec{c}_{\,n}$, where $N_k$ is the dimension of the kernel and $\mu^n$ are $N_k$ coefficients depending on reaction rates. Under the $j$-th knockdown, we have 
 \bal
 \delta_{j} \vec W   =\sum_{n=1}^{N_k}\delta_j \mu^n  \vec{c}_{\,n} = \biggl( \sum_{n=1}^{N_k}\, \frac{d \mu^n }{d k_j} \,  \vec{c}_{\,n}\biggr) \delta k_j.
 \label{dw1}
 \eal
The $i$-th component of $\delta_j \vec W$ is  also expanded as
\bal
\delta_j W_i =\biggl( \frac{\partial W_i}{\partial k_j} + \sum_{m'=1}^M \frac{\partial W_i }{\partial x_{m'}}\frac{d x_{m'}}{d k_j}\biggr)\delta k_j . \label{dw2}
\eal


From  \eqref{dw1}   \eqref{dw2}, the  response of steady state concentration $ \delta_j \vec x \equiv \frac{d \vec x}{d  k_j} \delta k_j $ \textcolor{black}{(circles in   FIG.    \ref{fig:NW})} and  flux $\delta_j \vec W$ \textcolor{black}{(arrows in   FIG.    \ref{fig:NW})} to each perturbation  $k_j \rightarrow k_j+ \delta k_j$ is determined from network structure only \cite{Mochizuki, monomolecular}. The result for each perturbation is simultaneously obtained through the following matrix computation:
\begin{eqnarray}
 \left(
\begin{array}{ccc}
\delta_1 {  \vec x} & \delta_2 { \vec   x}& \ldots \delta_R { \vec  x}\\\hline
\delta_1 \vec  \mu & \delta_2  \vec \mu & \ldots  \delta_R\vec  \mu 
\end{array}
\right)
\propto {\bf A}^{-1}\equiv {\bf S}\label{Sense}
\end{eqnarray}
where  the  matrix $\bf A$ is given as 
\begin{eqnarray}
{\bf A} = 
\left(
\begin{array}{cccc|c }
&\mbox{\smash{\large $\frac{\partial W_i}{\partial x_m}$}}&&&- {\vec c}_{\,1}\ \ldots\ - {\vec c}_{\,N_{k}}
\end{array}
\right).\label{Amatrix}
\end{eqnarray}
In \eqref{Sense} and \eqref{Amatrix}, the horizontal and vertical lines are the partitions of  the matrices. We then obtain the flux change $\delta_j \vec W$ from \eqref{dw1},  or 
\begin{align}
\left(\begin{array}{c  }\delta_{1} {\vec W}\ldots \delta_{R} {\vec W} \end{array}\right)=
\left(\begin{array}{c  }{\vec c}_{\,1} \ldots  { {\vec c}_{\,N_{k}}}\end{array}\right)\left(\begin{array}{ c }\delta_1 \vec  \mu \ldots \delta_R  \vec  \mu\end{array}\right)\label{fluxresponse}
\end{align}
in a matrix notation. 
 We call the inverse of $\bf A$ as the {\it sensitivity matrix}  $\bf S$. 
Note that  $\delta_j \vec x$, $\delta_j \vec \mu$, $\delta_j \vec W$,  $\vec c_n$ are column vectors with 
$M, N_k, R$, and $R$ components respectively, and $\frac{\partial W_i}{\partial x_m}$ is  an
 $R$-by-$M$ matrix.  We assume networks with ${\rm dim\, ker}\,  {\bm \nu}^T =0$ throughout this paper, which guarantees  the matrix $\bf A$ is square, i.e. $R=M+N_k$. 

\textcolor{black}{Comments are in order.  First, our theory depends only on the structure of reaction networks. The network structure is reflected in the distribution of nonzero entries in the $\bf A$-matrix, which  determines the   qualitative responses. Second, as a generalization of our method,  we  account for regulations such as allosteric effects by relaxing  \eqref{noreg} as
\begin{align}
\begin{cases}
\frac{\partial W_i}{\partial x_m}     \neq 0 \  \    {\rm  if}\  x_m \ {\rm influences \ reaction} \ i,  \\
  \frac{\partial W_i}{\partial x_m} =0\  \  {\rm otherwise.}\ 
  \end{cases} 
  {\tag{2'}}
\end{align}
Then, regulations  add  additional $r_{im}$ in the $\bf A$-matrix, and  the response is still determined through \eqref{Sense}.}

\section{Localization and Hierarchy}
Let us see some results of structural sensitivity analysis.

\begin{figure}[h] 
  \includegraphics[width=6cm,bb=0 20 300 160]{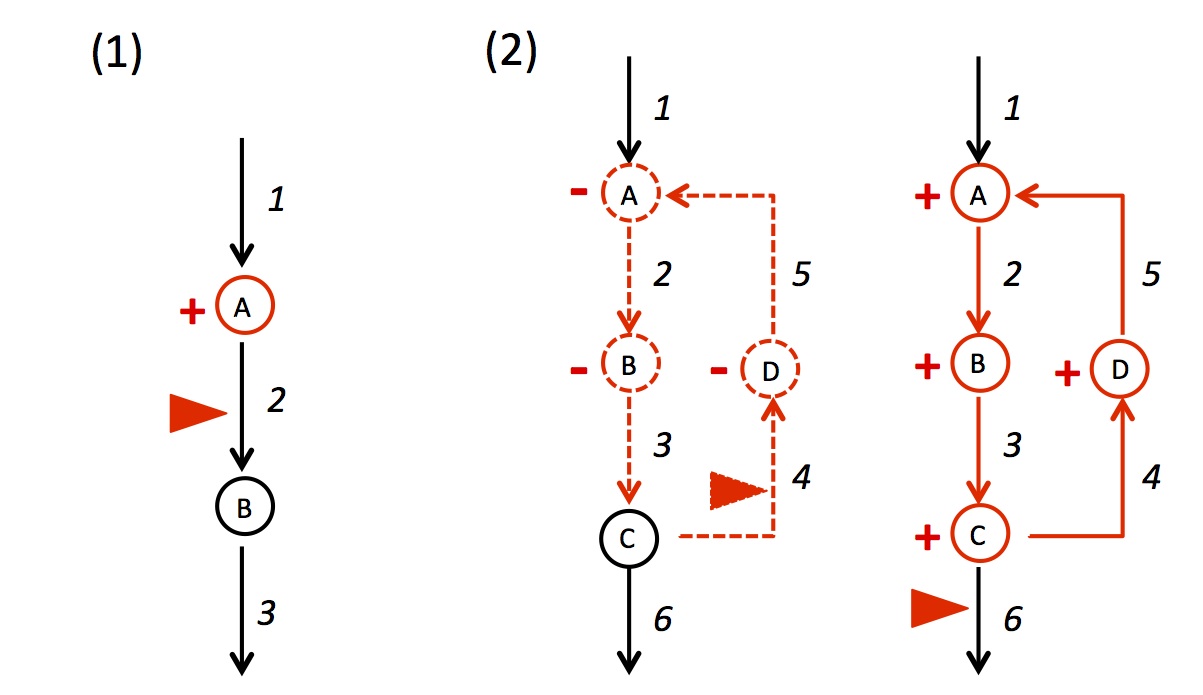}
   \caption{ Reaction networks and sensitivities in Example 1 and 2.   The  red triangle indicates a knocked down reaction. The signs (increase/decrease) of responses are represented by  $+/-$ for chemicals and solid/dashed red lines  for fluxes. }
   \label{fig:NW}
\end{figure}

$Example\ 1$: We consider a straight pathway, shown in FIG. \ref{fig:NW} (Left). The $\A$-matrix and the sensitivity matrix $\bf S$ are
\bal
\A = \left(
    \begin{array}{cc|c}
0 & 0 & -1\\
r_{2A}& 0 & -1\\
0 & r_{3B}&-1
\end{array}
  \right), 
{\bf S} = 
\left(
    \begin{array}{ccc}
- r_{2A}^{-1} & r_{2A}^{-1} & 0\\
-r_{3B}^{-1}& 0 & r_{3B}^{-1} \\\hline
-1 &0&0
\end{array}
  \right).\label{Sense1}
\eal
The  flux changes only when we perturb the top reaction 1 (the 1st column of $\bf S$). 
The perturbation to  reactions 2 or 3  changes only its substrate concentration  (the 2nd, 3rd column of $\bf S$). 

$Example\ 2$:
The second example  shown  in FIG. \ref{fig:NW} (Right) consists of 6 reactions and 4 chemicals. 
 The matrices {\bf A} and $\bf S$ are
\bal
\scalebox{0.8}{$ \A = \left(
    \begin{array}{cccc|cc}
0 & 0 & 0 & 0 & -1& 0 \\
r_{2A} & 0 & 0 & 0 & -1& -1\\
0 & r_{3B} & 0 & 0 & -1& -1\\
0 & 0 & r_{4C} & 0 & 0 & -1\\
0 & 0 & 0 & r_{5D} & 0 & -1\\
0 & 0 & r_{6C} & 0 & -1 & 0\\
\end{array}
  \right), 
  $}
\eal
\bal
\scalebox{0.8}{$ 
{\bf S} = 
\left(
    \begin{array}{cccccc}
\frac{-r_{4C}-r_{6C}}{r_{2A}r_{6C}} & r_{2A}^{-1} & 0 & - r_{2A}^{-1} & 0& \frac{r_{4C}}{r_{2A}r_{6C}} \\
\frac{-r_{4C}-r_{6C}}{r_{3B}r_{6C}} & 0 & r_{3B}^{-1} & -r_{3B}^{-1} & 0 &  \frac{r_{4C}}{r_{3B}r_{6C}} \\
- \frac{1}{r_{6C}}& 0 & 0 & 0 & 0&  r_{6C}^{-1} \\ 
 -\frac{r_{4C}}{r_{5D}r_{6C}} & 0 & 0 & -r_{5D}^{-1} &r_{5D}^{-1} &  \frac{r_{4C}}{r_{5D}r_{6C}} \\ \hline
-1 & 0 & 0 & 0 & 0& 0 \\
- \frac{r_{4C}}{r_{6C}} & 0 & 0 & -1 & 0& \frac{r_{4C}}{r_{6C}}.
\end{array}
  \right).$}\label{Sense2}
\eal

Again, only the perturbation to the input rate, corresponding to the 1st column in \eqref{Sense2},  affect all chemicals and fluxes. Perturbations to reactions $2, 3, 5$ only decrease the concentrations of the substrates $A, B, D$ respectively. 
Knockdown of reaction $4$ decreases the  concentrations $D, A, B$ along the  cycle downward of the perturbation (see FIG. \ref{fig:NW}, and the 4th column of $\bf S$). Knockdown of reaction 6 does not change  the further downstream but does change $A,B,C,D$ in the  cycle.  Also, the signs  of the responses are reversed (the 6th column of $\bf S$).  

\begin{figure}[htbp] 
  \includegraphics[width=3.cm,bb=120 0 350 330]{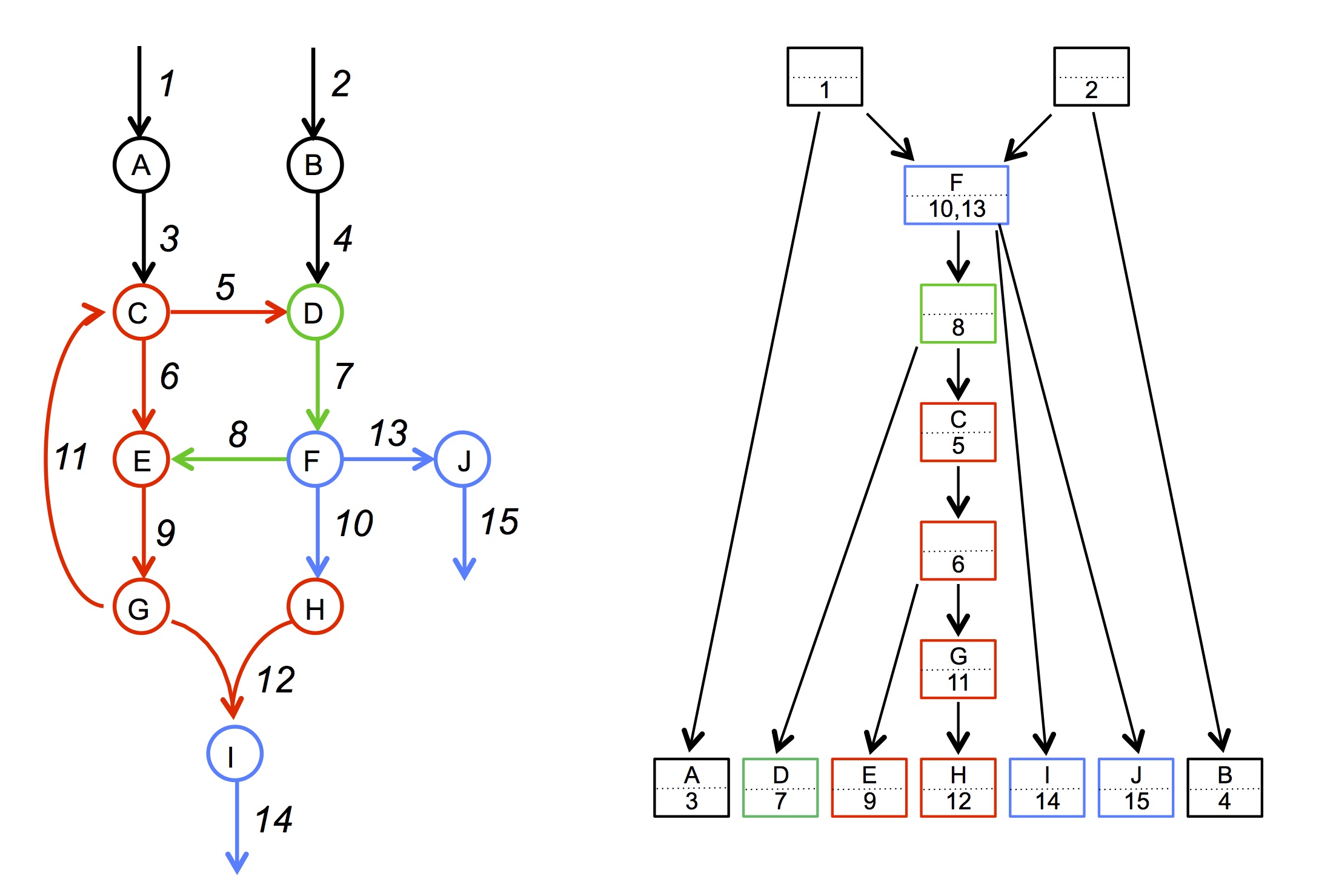}
  \vspace{-0.2cm}
   \caption{ (Left) Reaction network of example 3.  (Right) Graph of response hierarchy, summarizing the inclusion relations between nonzero response patterns. When a reaction rate in any square box is perturbed, the metabolites in the box plus those in the lower boxes exhibit nonzero responses.  The three colors (red, green, blue) correspond to $\Gamma_{10},\Gamma_{11},\Gamma_{12}$ respectively in the text.}
  \label{fig:NW3}
\vspace{-0.3cm}
\end{figure}

 $Example \ 3$: The third network in FIG. \ref{fig:NW3} (Left) includes 10 chemicals and 15 reactions.  FIG. \ref{fig:NW3} (Right) shows nonzero response patterns of metabolites and inclusion relation between them. See Appendix for the $\A$-matrix and the sensitivity matrix. S

In general, response to perturbations in chemical reaction networks exhibits two characteristics, $localization$ and $hierarchy$. The $localization$ means that  the influence of the perturbations is  confined in a finite region in a network. In other words, the naive intuition that a perturbation in an upper part of a reaction network influences all of the lower parts is incorrect.
  The $hierarchy$ implies that the nonzero response patterns under  perturbations of different reaction rates exhibit inclusion relations among them. 


\section{The Law of Localization}
From  the $\A$-matrix \eqref{Amatrix}, we can generally prove a   theorem, the {\it law of localization}, that determines the extent to which a perturbation influences in a network. 
\textcolor{black}{ For a given network, we consider a pair $\Gamma=({\m}, {\r})$ of a metabolite subset $\m$ and a reaction subset $\r$ satisfying the condition that $\r$ includes all reactions influenced by metabolites in $\m$ (see the condition (2')). The choice of $\r$ for a chosen $\m$ is not unique in general.} We call a subnetwork satisfying this condition ``output-complete.'' For such a subnetwork $\Gamma$, we count the number $|\m|$ of elements in $\m$, the number $|\r|$ of elements in $\r$, and the number $N_k(\r)$ of the closed cycles that consist of the reaction subset $\r$. Then, we compute an  index, 
\begin{equation}
\lambda(\Gamma)\equiv  - |{\m}|+ |{\r} |- N_k({\r}), \label{LoL}
\end{equation}
which is analogous to Euler characteristic and generally non-negative. 
The {\it law of localization} states that if $\lambda(\Gamma)=0$ for an output-complete subnetwork $\Gamma$, then any perturbation of reactions in $\Gamma$ does not change the  concentrations and the fluxes outside of $\Gamma$, namely the perturbation effect is localized in $\Gamma$ itself. We call an output-complete subnetwork satisfying $\lambda(\Gamma)=0$   {\it buffering structure}.

{\it Proof.--} The  theorem is proved  from the distribution of nonzero entries of the $\bf A$-matrix. (i) Suppose a subnetwork $\G$ is a buffering structure. 
Then by appropriately choosing a basis of the kernel of ${\bm \nu}$ and the orderings of the indices of the $\bf A$-matrix, we can always rewrite the  $\bf A$-matrix as
\bal
\scalebox{0.9}{$
{\bf A}=
 \begin{array}{c }
 \ _{ |{\r}|}\Bigg{\updownarrow} \\ 
\\
\end{array}
\overset{\overset{\ \ |{\m}|+N_k({\r})}{\quad \xlongleftrightarrow[]{\hspace{2.5em}}}\quad \quad \quad \quad }{\left(
\begin{array}{ccc|c }
&&&  \\
&\underset{square}{\mbox{\smash{\Large $*$}}}&&\ \ \ \ \mbox{\smash{\Large $*$}}\ \ \  \\
&&& \\ \hline
&&  &\\ 
&\mbox{\smash{\Large $\bf 0$}}&& \ \ \ \mbox{\smash{\Large $*$}}\ \ \  
\end{array}
\right)}.
$}
\label{Agamma}
\eal
The lower left block vanishes because $\bf A$  is output-complete.
(ii) As explained already, the concentration change $\delta_j x_m $ is proportional to ${\bf A}^{-1}_{mj} \propto  {\rm Det} \ \hat{{\bf A}}^{(j;m)}$, where $\hat{{\bf A}}^{(j;m)}$ is the minor matrix associated with the row of the $j$-th reaction and the column of the $m$-th metabolite. Then, ${\rm Det}\  \hat{{\bf A}}^{(j;m)}=0$ for $i \in \r $, $m \notin \m$  follows because  the upper left block in the minor $\hat{{\bf A}}^{(j;m)}$, which was originally square in \eqref{Agamma}, is horizontally long. \hspace{\fill}$\square$


We illustrate the {\it law of localization} in the example networks in FIG. \ref{fig:NW} and FIG. \ref{fig:NW3} (Left).

 $Example \ 1$: The network includes two buffering structures, $\G_1=(\{A\},\{2\})$ and $\G_2=(\{B\},\{3\})$ which are minimum buffering structures including only a single chemical and a single output reaction. The {\it law of localization} claims that the perturbation to reaction 2 in $ \Gamma_1$  influence only the inside of $\G_1$, namely the concentration of A. Note  the flux 2 in $\Gamma_1$ does not change in order to keep the outside of $\Gamma_1$ unchanged). We actually observed the predicted response in \eqref{Sense1}. Generally, a perturbation to a reaction which is a single output  from a chemical  influences the substrate concentration only. 

 $Example \ 2$: In addition to the 3 minimal buffering structures, $\G_1=(\{A\},\{2\}),\  \G_2=(\{B\},\{3\}),\ \G_3=(\{D\},\{5\})$, the network has two larger ones, $\G_4=(\{A,B,D\},\{2,3,4,5\})$ (with $\lambda(\G_4)=-3+4-1=0$), $\G_5=(\{A,B,C,D\},\{2,3,4,5,6\})$ (with $\lambda(\G_5)=-4+5-1=0$). $\G_4$ is the minimum buffering structure including reaction 4. Then,  the {\it law of localization} predicts that the nonzero response to perturbation of reaction 4 should be limited within $\G_4$, which is observed in the 4th column in \eqref{Sense2}. Similarly, the response to perturbation of reaction 6 is explained by $\G_5$. 

 $Example \ 3$: The network  has 14 buffering structures,  listed in Appendix. To examine the  response hierarchy, we focus on the three buffering structures colored in FIG. \ref{fig:NW3}; $\G_{10}=(\{C,E,G,H\},\{5,6,9,11,12\})$ (with $\lambda(\G_{10})=-4+5-1=0$), $\G_{11}=(\{C,D,E,G,H\},$$\{5,6,7,8,9,11,12\})$ (with $\lambda(\G_{11})$$=-5+7-2=0$), and $\G_{12}=(\{ C,D,E,F,G,H,I,J\},\{5,6,7,8,9,10,11,12,13,14,15\})$ (with $\lambda(\G_{12})$$=-8+11-3=0$).  Each of these three explains the response pattern under perturbation of reaction 5, 8, and 10 (or 13), respectively, and they  satisfy an inclusion relation, $\G_{10} \subset \G_{11} \subset \G_{12}$.  Accordingly,  we can see from FIG. \ref{fig:NW3} (Right) that these response patterns satisfy an inclusion relation.


In this way,    we  understand all of the observed patterns  from network topology by using the {\it law of localization}. In short, the first characteristic, $localization$, is explained from the existence of buffering structures. The second property, $hierarchy$, is  explained as the nest of the buffering structures. 

\begin{figure}[htbp]
  \includegraphics[width=8cm,bb=0 20 205 225]{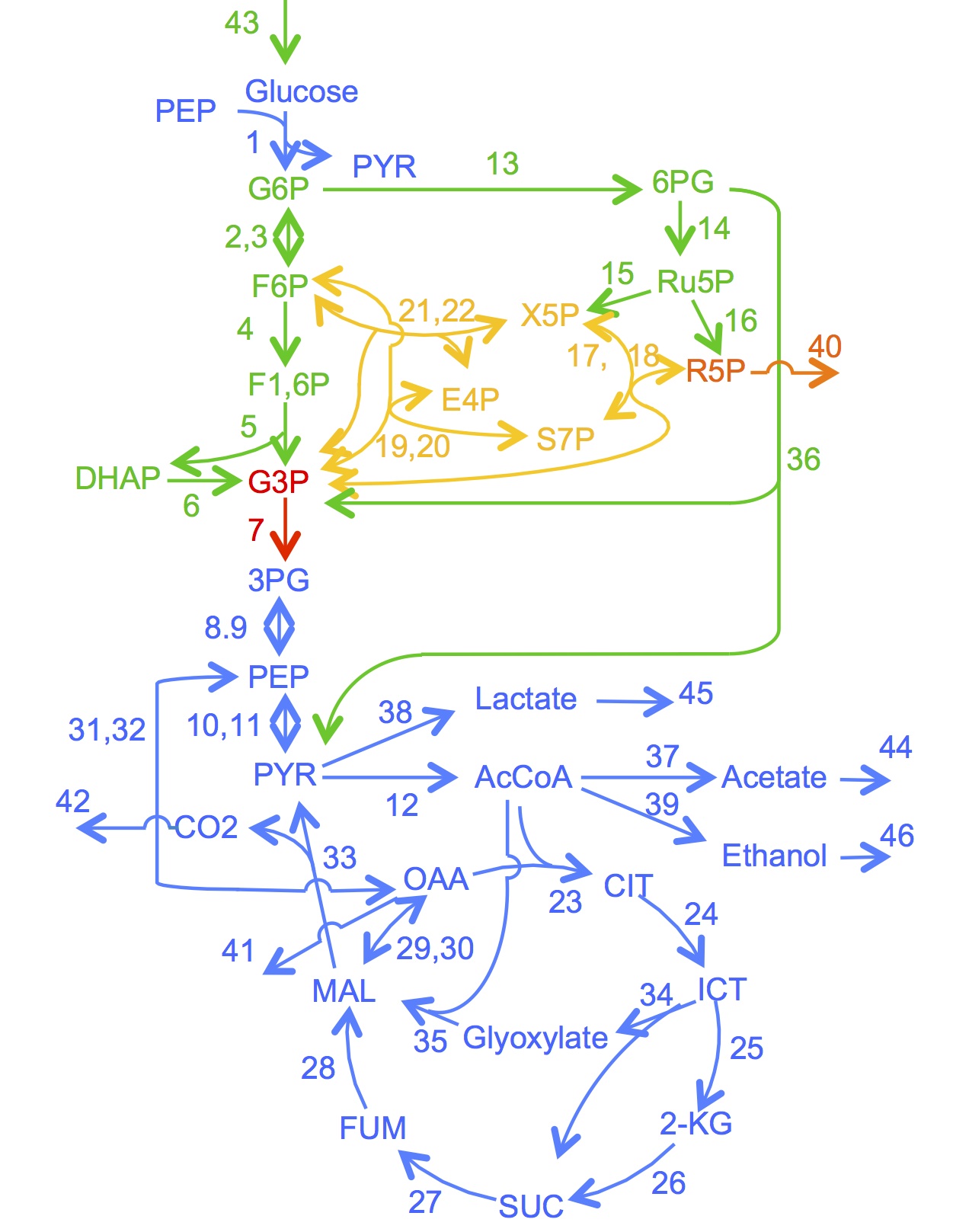}
   \caption{ E. coli network. (Adopted from \cite{Ishii}). 
  }
   \label{fig:zu1}
\end{figure}
\begin{figure}[htbp]
  \includegraphics[width=14.cm,bb=10 10 400 100]{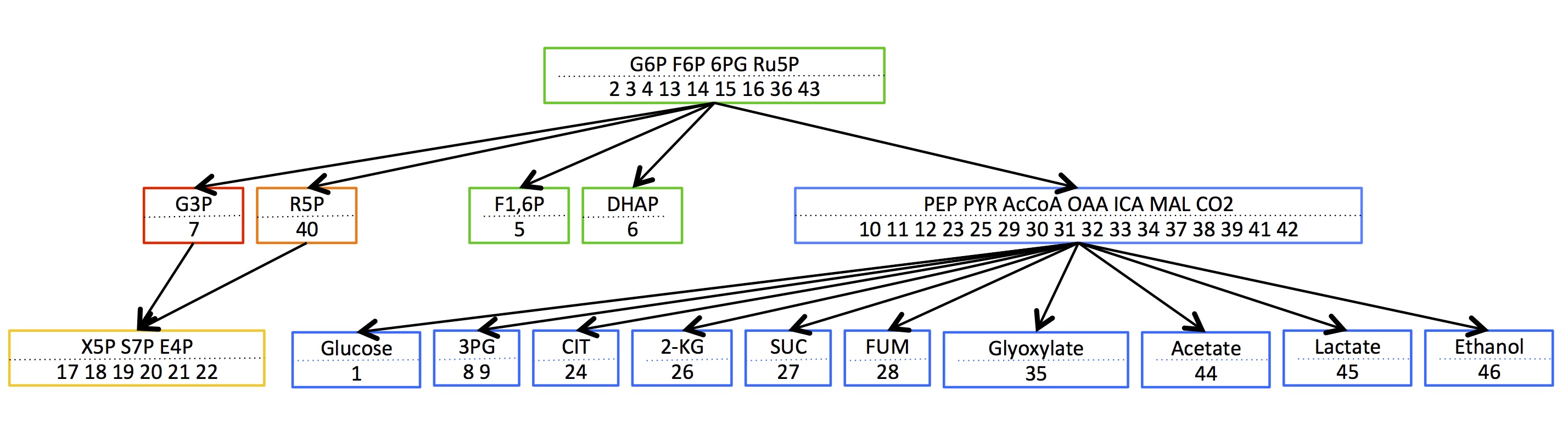}
   \caption{  The response hierarchy of E. coli network. 
  }
   \label{fig:zu2}
\end{figure}

Finally, as an application  to real biological networks, we examine the carbon metabolism pathway of $E.\ coli$. The network  is a major part of the energy acquisition process, and the basic structures  are shared between bacteria and human beings. 
FIG. \ref{fig:zu1} shows the network \cite{Ishii}, including 28 metabolites and 46 reactions, and FIG . \ref{fig:zu2} shows the response hierarchy (see Appendix for the detail). Again, the response patterns exhibit the two characteristic features, $localization$ and $hierarchy$. The network has 17 buffering structures, and the existence and the nest of them explain the two characteristic features perfectly.  
We mention that some of the buffering structures, which are of course defined from network topology,  surprisingly overlap biologically identified sub-circuits, the pentose phosphate pathway (yellow in FIG. \ref{fig:zu1}, \ref{fig:zu2}), the tricarboxylic acid cycle (blue) and the glycolysis (green). This correspondence may be understood from an evolutional point of view by considering the advantage of buffering structures.




\section{Discussions and Conclusions}
Here we discuss the biological significances of buffering structures (and nest of them) in two different levels. The first discussion is on the physiological importance. A buffering structure 
 prohibits influence of given perturbation from expanding to  the outside, like a ``firewall.'' In other words, it is a substructure with robustness emerging from the network topology. The carbon metabolism network of E. coli possesses multiple nested firewalls (FIG. \ref{fig:zu2}), and are expected to be robust to fluctuations of enzymes in it. We   expect that such a topological characteristic of reaction networks could be the evolutionary origin of homeostasis in biological systems. A set of chemical reactions satisfying  the condition of buffering structure by chance in evolutionarily early time  would be positively selected as an advantageous circuit. We then expect that buffering structures  in existing biological networks today  might be generated and selected in such ways. 

The second discussion is about practicality of the {\it law of localization} in  experimental biology. Our knowledge of biochemical  networks is considered incomplete: There might exist unidentified reactions or regulations. 
 The condition for buffering structure depends on the local network structure only, which implies that we can study the sensitivity of the system  only from local information on the network.

\begin{figure}[htbp]
  \centering 
  \includegraphics[width=12cm,bb=0 20 550 150]{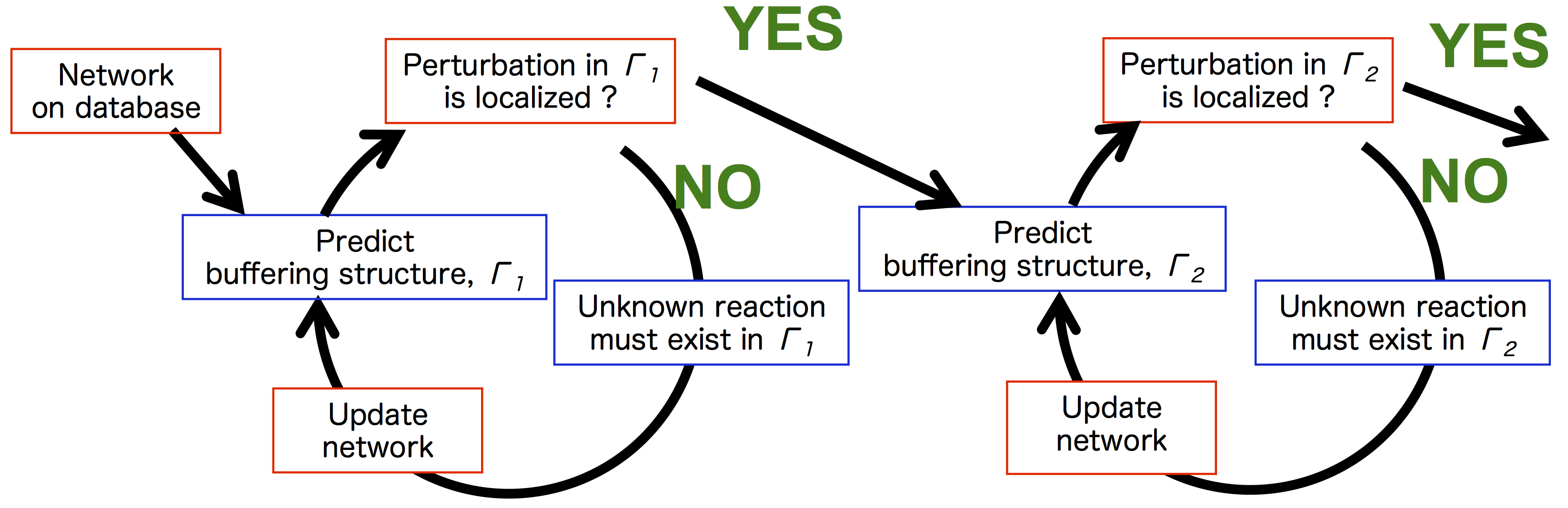}
   \caption{A strategy toward elucidating a true network. }
   \label{fig:strategy}
   \vspace{-0.2cm}
\end{figure}

From this property, we can  determine a ``true'' network by combining experiments as shown in FIG. \ref{fig:strategy}. 
  If a given perturbation (knock down or overexpression) to a predicted buffering structure, determined from network topology, $does$ influence outside of the buffering structure, then there must be inconsistency between the database information and the actual network. The mismatch must exist inside of the candidate structure, i.e. there must be unknown reactions or unknown regulations inside (or emanating from) the candidate subnetwork.  By repeating theoretical predictions and experimental verifications, we can determine  the ``true'' network   from a partial network  to the whole network in a step-by-step manner, i.e. from smaller to larger buffering structures. Our theory must promote the understanding of reaction networks in both the theoretical and experimental levels by directly connecting  the network topology with behaviors of the systems.

 Using a different method, Steuer {\it et al.} studied a mathematical criteria for ``perfect adaptation,'' where changing a rate constant in one part of the network does not affect steady-state concentrations or fluxes, which in fact, is a subpart of the phenomena we studied in this paper. There are at least three large differences: (i) We studied not only perfect adaptation, but also any qualitative responses (increase/decrease/invariant), (ii) While Steuer et al.'s method needs to examine a  condition one by one for each pair of  perturbation and  chemicals, our method determines changes of all concentrations and fluxes by each perturbation of all reaction rates simultaneously via \eqref{Sense}. (iii) We found and proved a general law which claims that the property of perfect adaptation emerges from local topology of network. Despite these differences, it would be interesting to explore relations between two mathematical theories.

This work was supported partly by the CREST, Japan Science and Technology Agency, and by iTHES research program RIKEN, by Grant-in-Aid for Scientific Research on Innovative Area, ``Logics of Plant Development,"  Grant number 25113005.  We greatly appreciate  Bernold Fiedler, Hiroshi Matano, and Hannes Stuke for their mathematical discussions. We also express our sincere thanks to   Testuo Hatsuda, Michio Hiroshima, Yoh Iwasa, Sinya Kuroda,  Masaki Matsumoto, Keiichi Nakayama, Madan Rao, and Yasushi Sako  for their helpful discussions and comments.

\nocite{*}


\appendix

\section{The analysis of Example 3}
\subsubsection{The $\bf A$ matrix and sensitivity matrix $\bf S$ for Example 3}
The $\bf A$ matrix for the network of Example 3 in the main text is given by 
\bal
{\bf A}= \left(
\begin{array}{cccccccccc|ccccc}
 0 & 0 & 0 & 0 & 0 & 0 & 0 & 0 & 0 & 0 & 0 & 0 & 0 & 0 & -1 \\
 0 & 0 & 0 & 0 & 0 & 0 & 0 & 0 & 0 & 0 & -1 & 0 & 0 & 0 & 1 \\
 r_{3 A} & 0 & 0 & 0 & 0 & 0 & 0 & 0 & 0 & 0 & 0 & 0 & 0 & 0 & -1 \\
 0 & r_{4 B} & 0 & 0 & 0 & 0 & 0 & 0 & 0 & 0 & -1 & 0 & 0 & 0 & 1 \\
 0 & 0 & r_{5 C} & 0 & 0 & 0 & 0 & 0 & 0 & 0 & 0 & 1 & 0 & -1 & -1 \\
 0 & 0 & r_{6 C} & 0 & 0 & 0 & 0 & 0 & 0 & 0 & 0 & -1 & -1 & 1 & 0 \\
 0 & 0 & 0 & r_{7 D} & 0 & 0 & 0 & 0 & 0 & 0 & -1 & 1 & 0 & -1 & 0 \\
 0 & 0 & 0 & 0 & 0 & r_{8 F} & 0 & 0 & 0 & 0 & 0 & 0 & 0 & -1 & 0 \\
 0 & 0 & 0 & 0 & r_{9 E} & 0 & 0 & 0 & 0 & 0 & 0 & -1 & -1 & 0 & 0 \\
 0 & 0 & 0 & 0 & 0 & r_{10 F} & 0 & 0 & 0 & 0 & 0 & -1 & 0 & 0 & 0 \\
 0 & 0 & 0 & 0 & 0 & 0 & r_{11 G} & 0 & 0 & 0 & 0 & 0 & -1 & 0 & 0 \\
 0 & 0 & 0 & 0 & 0 & 0 & r_{12 G} & r_{12 H} & 0 & 0 & 0 & -1 & 0 & 0 & 0 \\
 0 & 0 & 0 & 0 & 0 & r_{13 F} & 0 & 0 & 0 & 0 & -1 & 2 & 0 & 0 & 0 \\
 0 & 0 & 0 & 0 & 0 & 0 & 0 & 0 & r_{14 I} & 0 & 0 & -1 & 0 & 0 & 0 \\
 0 & 0 & 0 & 0 & 0 & 0 & 0 & 0 & 0 & r_{15 J} & -1 & 2 & 0 & 0 & 0 \\
\end{array}
\right).\nonumber
\eal
Here the row indices are the reactions $1,\ldots,15$, and the column indices are \bal {A,B,C,D,E,F,G,H,I,J,{\bf c_1},{\bf c_2},{\bf c_3},{\bf c_4},{\bf c_5}},\eal
where $\{{\bf c}_i\}$ is a basis of the kernel space of the stoichiometric matrix $S$. The vertical line separates the indices of the chemicals and those of the kernel vectors.
By inverting  $\bf A$, we obtain the sensitivity matrix $\bf S$. The result is
${\bf S} = - D_1 \, {\bf s} \, D_2$, where $D_1, D_2$ are the diagonal matrices defined as  
 \bal
 D_1 &=diag (r_{3 A},r_{4 B},r_{5 C},r_{7 D},r_{5 C} r_{9 E},1,r_{5 C} r_{11 G},
 r_{5 C} r_{11 G} r_{12 H},r_{14 I},r_{15 J},1,1,r_{5 C},1,1)^{-1},\nonumber \\
 D_2&=diag(R_3,R_3,
 1,1,1,1,1,1,1,R_3,1,1,R_3,1,1)^{-1}, \nonumber
\eal
and  ${\bf s}$ is defined as 
{\scriptsize
$$
{\bf s}=\left(
\begin{array}{ccccccccccccccc}
 {R_3} & 0 & -1 & 0 & 0 & 0 & 0 & 0 & 0 & 0 & 0 & 0 & 0 & 0 & 0 \\
 
 0 & {R_3 }& 0 & -1 & 0 & 0 & 0 & 0 & 0 & 0 & 0 & 0 & 0 & 0 & 0 \\
 
 R_4 & R_2 & 0 & 0 & -1 & 0 & 0 & 1 & 0 & -R_5 & 0 & 0 & -R_2 & 0 & 0 \\
 
 R_4 & R_4 & 0 & 0 & 0 & 0 & -1 & 1 & 0 & -R_5 & 0 & 0 & -R_2 & 0 & 0 \\
 
{\bf s}_{E,1} & {\bf s}_{E,2} & 0 & 0 & -r_{6 C} & r_{5 C} & 0 & R_1 & -r_{5 C} &{\bf s}_{E,10}  & 0 & 0 &{\bf s}_{E,13}  & 0 & 0 \\
 1 & 1 & 0 & 0 & 0 & 0 & 0 & 0 & 0 & -2 & 0 & 0 & -1 & 0 & 0 \\
 {\bf s}_{G,1} & R_1 R_2 & 0 & 0 & -r_{6 C} & r_{5 C} & 0 & R_1 & 0 & -R_1 R_5 & -r_{5 C} & 0 & -R_1 R_2 & 0 & 0 \\
 {\bf s}_{H,1} & {\bf s}_{H,2} & 0 & 0 & r_{6 C} r_{12 G} & -r_{5 C} r_{12 G} & 0 & -R_1 r_{12 G} & 0 &  {\bf s}_{H,10}  & r_{5 C} r_{12 G} & -r_{5 C} r_{11 G} & {\bf s}_{H,13} & 0 & 0 \\
 
 r_{10 F} & r_{10 F} & 0 & 0 & 0 & 0 & 0 & 0 & 0 & r_{13 F} & 0 & 0 & -r_{10 F} & -1 & 0 \\
 
 r_{13 F} & r_{13 F} & 0 & 0 & 0 & 0 & 0 & 0 & 0 & -2 r_{13 F} & 0 & 0 & 2 r_{10 F} & 0 & -1\\
\hline
 R_3 &  R_3 & 0 & 0 & 0 & 0 & 0 & 0 & 0 & 0 & 0 & 0 & 0 & 0 & 0 \\
 r_{10 F} & r_{10 F} & 0 & 0 & 0 & 0 & 0 & 0 & 0 & r_{13 F} & 0 & 0 & -r_{10 F} & 0 & 0 \\
 {\bf s}_{{\bf c}_3,1} & R_1 R_2 & 0 & 0 & -r_{6 C} & r_{5 C} & 0 & R_1 & 0 & -R_1 R_5 & 0 & 0 & -R_1 R_2 & 0 & 0 \\
 r_{8 F} & r_{8 F} & 0 & 0 & 0 & 0 & 0 & 1 & 0 & -2 r_{8 F} & 0 & 0 & -r_{8 F} & 0 & 0 \\
 R_3 & 0 & 0 & 0 & 0 & 0 & 0 & 0 & 0 & 0 & 0 & 0 & 0 & 0 & 0 \\
\end{array}
\right).\nonumber
$$
}
Here we have defined  the factors $R_i$($i=1,\ldots,5$)  by
\bal
R_1&=r_{5C}+r_{6C}, \nonumber\\
R_2&=r_{8F}-r_{10F}, \nonumber\\
R_3 &= 2 r_{10 F} + r_{13 F}, \nonumber\\
R_4 &= r_{8F} +r_{10F}+r_{13F}, \nonumber\\
R_5 &= 2 r_{8F}+r_{13F}, \nonumber
\eal
and the components ${\bf s}_{m,i}$  in $\bf s$  by 
\bal 
{\bf s}_{E,1}&=R_4 r_{6 C}+r_{5 C} r_{8 F},\nonumber\\
{\bf s}_{E,2}&=R_1 r_{8 F}-r_{6 C} r_{10 F},\nonumber\\ 
{\bf s}_{E,10}&= -2 R_1 r_{8 F}-r_{6 C} r_{13 F} , \nonumber\\
{\bf s}_{E,13} &=r_{6 C} r_{10 F}-R_1 r_{8 F}, \nonumber\\
{\bf s}_{G,1}&=R_2 r_{5 C}+R_4 r_{6 C} ,\nonumber\\
 {\bf s}_{H,1} &=r_{5 C} \left(r_{10 F} r_{11 G}-R_2 r_{12 G}\right)-R_4 r_{6 C} r_{12 G} , \nonumber\\
 {\bf s}_{H,2} &= r_{5 C} r_{10 F} r_{11 G}-R_1 R_2 r_{12 G}, \nonumber\\
 {\bf s}_{H,10}& =r_{5 C} r_{13 F} r_{11 G}+R_1 R_5 r_{12 G} , \nonumber\\
 {\bf s}_{H,13} &=R_1 R_2 r_{12 G}-r_{5 C} r_{10 F} r_{11 G},\nonumber\\
 {\bf s}_{{\bf c}_3,1} &=R_2 r_{5 C}+R_4 r_{6 C}.\nonumber
\eal

\newpage 
\subsubsection{List of buffering structures}
The network of Example 3 has the following fourteen buffering structures (and unions of them).
\bal
\Gamma_1 &= (\{A\},\{3\}),\nonumber\\
\Gamma_2 &= (\{B\},\{4\}),\nonumber\\
\Gamma_3 &= (\{D\},\{7\}),\nonumber\\
\Gamma_4 &= (\{E\},\{9\}),\nonumber\\
\Gamma_5 &= (\{H\},\{12\}),\nonumber\\
\Gamma_6 &= (\{I\},\{14\}),\nonumber\\
\Gamma_7 &= (\{J \},\{15\}),\nonumber\\
\Gamma_8 &= (\{G,H\},\{11,12\}),\nonumber\\
\Gamma_9 &= (\{E,G,H\},\{6,9,11,12\}),\nonumber\\
\Gamma_{10} &= (\{C,E,G,H\},\{5,6,9,11,12\}),\nonumber\\
\Gamma_{11} &= (\{C,D,E,G,H\},\{5,6,7,8,9,11,12\}),\nonumber\\
 \Gamma_{12} &=(\{C,D,E,F,G,H,I,J\},\nonumber\\
&\{ 5, 6, 7, 8, 9, 10, 11, 12, 13, 14, 15\} ),\nonumber\\
 \Gamma_{13} &=(\{A,C,D,E,F,G,H,I,J\},\nonumber\\
&\{1, 3, 5, 6, 7, 8, 9,10,11,12,13,14,15\}) ,\nonumber\\
\Gamma_{14} &=(\{B,C,D,E,F,G,H,I,J\},\nonumber\\
&\{2, 4, 5, 6, 7, 8, 9,10,11,12,13,14,15\}).\nonumber
\eal

\section{ E. coli  central metabolism}\label{app:ecoli}
\subsubsection{{ List of reactions}}\ 
1: Glucose  +  PEP  $\rightarrow$  G6P  +  PYR. 
 
 2: G6P   $\leftarrow$  F6P. 
 
 3: F6P   $\rightarrow$  G6P.
  
  4: F6P  $\rightarrow$  F1,6P. 
   
   5: F1,6P  $\rightarrow$  G3P  +  DHAP. 
   
   6: DHAP   $\rightarrow$  G3P.
   
 7: G3P   $\rightarrow$  3PG. 
  
  8: 3PG   $\rightarrow$  PEP. 
  
  9: PEP   $\rightarrow$  3PG.  
  
  10: PEP   $\rightarrow$  PYR. 
  
  11: PYR   $\rightarrow$  PEP. 
  
  12: PYR    $\rightarrow$  AcCoA  +   CO2. 
  
  13: G6P  $\rightarrow$  6PG. 
  
  14: 6PG  $\rightarrow$   Ru5P  +  CO2. 
  
  15: Ru5P  $\rightarrow$  X5P.
  
   16: Ru5P  $\rightarrow$   R5P. 
   
   17: X5P  +  R5P  $\rightarrow$   G3P  +  S7P. 
   
   18: G3P  +  S7P  $\rightarrow$   X5P  +  R5P. 
   
   19: G3P  +  S7P  $\rightarrow$   F6P  +  E4P.
   
    20: F6P  +  E4P  $\rightarrow$  G3P  +  S7P. 
    
21:  X5P  +  E4P  $\rightarrow$   F6P  +  G3P. 
  
  22: F6P   +  G3P  $\rightarrow$   X5P  +  E4P.  
  
  23: AcCoA  +    $\rightarrow$  CIT. 
  
  24: CIT   $\rightarrow$  ICT. 
  
  25: ICT  $\rightarrow$  2${\rm \mathchar`-}$KG  +  CO2. 
  
  26: 2-KG  $\rightarrow$   SUC  +  CO2.
  
   27: SUC   $\rightarrow$  FUM. 
   
  28:  FUM  $\rightarrow$  MAL. 
   
   29: MAL   $\rightarrow$  OAA.
   
 30: OAA   $\rightarrow$  MAL.

 31: PEP  +  CO2  $\rightarrow$  OAA.

 32: OAA  $\rightarrow$  PEP  +   CO2. 

 33: MAL  $\rightarrow$   PYR  +  CO2.

34: ICT   $\rightarrow$  SUC  +  Glyoxylate. 

 35: Glyoxylate  +  AcCoA  $\rightarrow$  MAL. 

 36: 6PG  $\rightarrow$   G3P  +  PYR. 

 37: AcCoA  $\rightarrow$   Acetate. 

38:  PYR  $\rightarrow$  Lactate. 

 39: AcCoA  $\rightarrow$  Ethanol. 

 40: R5P  $\rightarrow$ (output).

 41: OAA  $\rightarrow$ (output).

 42: CO2  $\rightarrow$ (output).

43:  (input) $\rightarrow$  Glucose. 

 44:  Acetate $\rightarrow$ (output).
 
  45: Lactate $\rightarrow$ (output).

46:  Ethanol $\rightarrow$ (output).

\subsubsection{{ List of buffering structures}} \
The E. coli network exhibits the following 17 different buffering structures $\G_i=(\m_i,\r_i) $  ($i=1,\ldots,17$).

{\scriptsize
$\Gamma_1=(\{ \rm 
Glucose
\},\{
1
\})$,

$\Gamma_2=(\{ \rm 
Glucose,
PEP,
G6P,
F6P,
F1,6P,
DHAP,
G3P,
3PG,
PYR,
6PG,
Ru5P,
X5P,
R5P, 
S7P,
E4P,
AcCoA,
OAA,
CIT,\\
ICT,
2{\rm \mathchar`-KG},
SUC,
FUM,
MAL,
CO2,
Glyoxylate,
Acetate,
Lactate,
Ethanol
\},  \{
1,
2 ,
3 ,
4 ,
5 ,
6 ,
7 ,
8 ,
9 ,
10 ,
11 ,
12 ,
13 ,
14 ,
15 ,\\
16 ,
17 ,
18 ,
19 ,
20 ,
21 ,
22 ,
23 ,
24 ,
25 ,
26 ,
27 , 
28 ,
29 ,
30 ,
31 ,
32 ,
33 ,
34 ,
35 ,
36 ,
37 ,
38 ,
39 ,
40 ,
41 ,
42 ,
44 ,
45 ,
46 
\})$,

$\Gamma_3=(\{ \rm 
F1,6P
\},\{
5 
\})$,

$\Gamma_4=(\{ \rm 
DHAP
\},\{
6 
\})$,

$\Gamma_5=(\{ \rm 
G3P,
X5P,
S7P,
E4P
\},\{
7 ,
17 ,
18 ,
19 ,
20 ,
21 ,
22 
\})$,

$\Gamma_6=(\{ \rm 
3PG
\},\{
8 
\})$,

$\Gamma_7=(\{ \rm 
Glucose,
PEP,
3PG,
PYR,
AcCoA,
OAA,
CIT,
ICT,
2{\rm \mathchar`-KG},
SUC,
FUM, 
MAL,
CO2,
Glyoxylate,
Acetate,
Lactate,
Ethanol
\},\\ \{
1,
8 ,
9 ,
10 ,
11 ,
12 ,
23 ,
24 ,
25 ,
26 ,
27 ,
28 ,
29 ,
30 ,
31 ,
32 ,
33 ,
34 ,
35 ,
37 ,
38 ,
39 ,
41 ,
42 ,
44 ,
45 ,
46 
\})$,

$\Gamma_8=(\{ \rm 
X5P,
S7P,
E4P
\},\{
17 ,
18 ,
19 ,
20 ,
21 
\})$ ,

$\Gamma_9=(\{ \rm 
CIT
\},\{
24 
\})$,

$\Gamma_{10}=(\{ \rm 
2{\rm \mathchar`-KG}
\},\{
26 
\})$,

$\Gamma_{11}=(\{ \rm 
SUC
\},\{
27 
\})$ ,

$\Gamma_{12}=(\{ \rm 
FUM
\},\{
28 
\})$ 

$\Gamma_{13}=(\{ \rm 
Glyoxylate
\},\{
35 
\})$,

$\Gamma_{14}=(\{ \rm 
X5P,
R5P,
S7P,
E4P
\},\{
17 ,
18 ,
19 ,
20 ,
21 ,
40 
\})$ ,

$\Gamma_{15}=(\{ \rm 
Acetate
\},\{
44 
\})$,

$\Gamma_{16}=(\{ \rm 
Lactate
\},\{
45 
\})$,

$\Gamma_{17}=(\{ \rm 
Ethanol
\},\{
46 
\})$.
}

\end{document}